\documentclass[letterpaper,superscriptaddress,twocolumn,showpacs,prl]{revtex4}%
\usepackage[dvips]{graphicx}
\usepackage{dcolumn}
\usepackage{bm}
\usepackage{amsmath}
\usepackage{color}
\usepackage{amsfonts}
\usepackage{amssymb}
\usepackage{subfigure}%
\providecommand{\U}[1]{\protect\rule{.1in}{.1in}}
\begin{document}
\author{HuJun Jiao}
\affiliation{Kavli Institute of NanoScience, Delft University of Technology, 2628 CJ Delft,
The Netherlands}
\author{Gerrit E. W. Bauer}
\affiliation{Institute for Materials Research, Tohoku University, Sendai 980-8577, Japan }
\affiliation{Kavli Institute of NanoScience, Delft University of Technology, 2628 CJ Delft,
The Netherlands}
\title{AC Voltage Generation by Spin Pumping and Inverse Spin Hall Effect }
\date{\today }

\begin{abstract}
The polarization of the spin current pumped by a precessing ferromagnet into
an adjacent normal metal has a constant component parallel to the precession
axis and a rotating one normal to the magnetization. The former component is
now routinely detected in the form of a DC voltage induced by the inverse spin
Hall effect (ISHE). Here we compute AC-ISHE voltages much larger than the DC
signals for various material combinations and discuss optimal conditions to
observe the effect. Including the backflow of spins is essential for
distilling parameters such as the spin Hall angle from ISHE-detected spin
pumping experiments.

\end{abstract}
\maketitle

In magnetoelectronics the electronic spin degree of freedom creates new
functionalities that lead to applications in information technologies such as
sensors and memories \cite{Maek06}. Central to much excitement in this field
is the spin Hall effect (SHE) \cite{Hirs99,Zhang00,Mura03,Sino04},
\textit{i.e.} the spin current induced normally to an applied charge current
in the presence of spin-orbit interaction, as discovered optically in
semiconductors \cite{Kato04,Wun05} and subsequently electrically in metals
\cite{Sait06,Valen06,Kim07}. Recently magnetization reversal by the SHE
induced spin transfer torque has been demonstrated \cite{Cornell}. The
generation of a voltage by a spin current injected into a paramagnetic metal,
the \textit{inverse} spin Hall effect (ISHE), can be employed to detect the
spin current due to spin-pumping \cite{Tser02a,Tser05,Gui11} by an adjacent
ferromagnet under ferromagnetic resonance (FMR) conditions
\cite{Sait06,Azeve05}. The ISHE has also been essential for the discovery of
the spin Seebeck effect \cite{Uchida08}.

In recent experiments, DC voltages induced by the ISHE have been measured in
many material combinations therey giving access to crucial parameters such as
the spin Hall angle \cite{Mose10a,Mose10b,Azeve11} and the spin mixing
conductance \cite{Czes11}, \textit{i.e.} the material parameter determining
the effectiveness of interface spin-transfer torques \cite{Tser05}. For
example, the magnitude and sign of the ISHE as parameterized by the spin Hall
angle has been determined for permalloy (Py)$|$N bilayers for different normal
metals N \cite{Mose10a,Mose10b}. An approximate scaling relation for the spin
pumping by numerous ferromagnets (F) has been discovered by comparing
different F$|$Pt bilayers as a function of excitation power \cite{Czes11}.
However, it is far from easy to derive quantitative information from ISHE
experiments \cite{Hu}. As reviewed by the Cornell collaboration \cite{Liu12},
several experimental pitfalls should be avoided. At the FMR, the DC ISHE
voltage is small, scaling quadratically with the cone angle of the precessing
magnetization. An important correction is caused by the back-diffusion
(\textquotedblleft back-flow\textquotedblright) of injected spins to the
interface, which effectively reduces the spin current injection \cite{Tser05}
and generates voltages normal to the interface \cite{Xuhui06,Cost06}. This
backflow has often been neglected in interpreting spin pumping experiments,
assuming that Pt, the metal of choice, can be treated like a perfect spin sink.

The spin current injected by FMR into a normal metal consists of a DC
component along the $z$-axis parallel to the effective field and an AC
component normal to it, \textit{i.e}. in the $xy$-plane (see Fig. 1). In this
Letter we analyze both AC and DC ISHE voltages by time-dependent spin
diffusion theory, where the former is generated between the edges of the
sample along the $z$-direction,\textit{ i.e.} for a different sample
configuration than used for DC signal detection. For small precession angles
the AC ISHE voltage is found to be orders of magnitude larger than the DC
signal. The back-flow of spins modifies also the DC voltage even for small
spin-flip diffusion lengths, requiring a reappraisal of published parameters.
\begin{figure}[ptb]
\includegraphics[width=3.2cm]{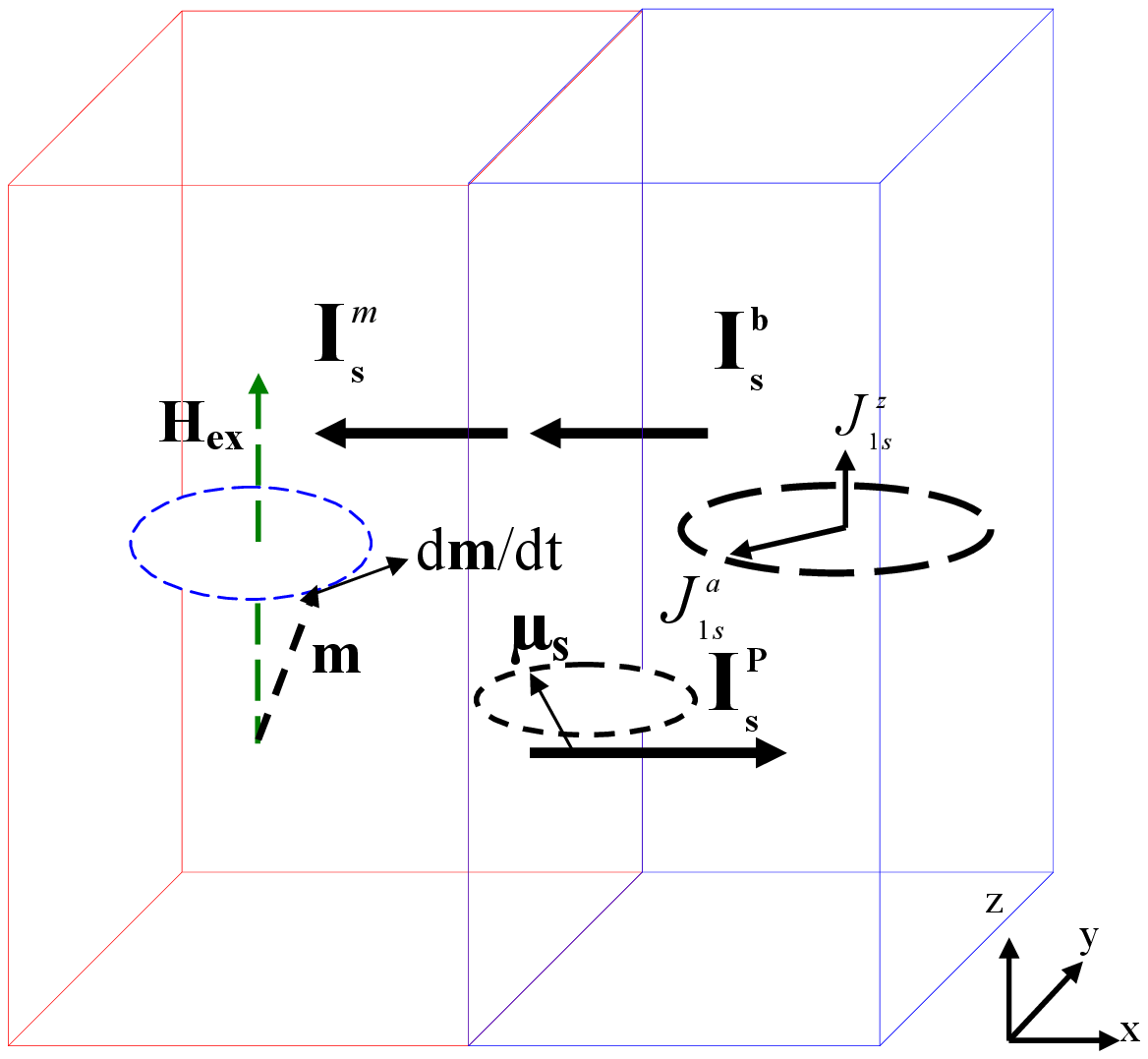}
\includegraphics[width=4.2cm]{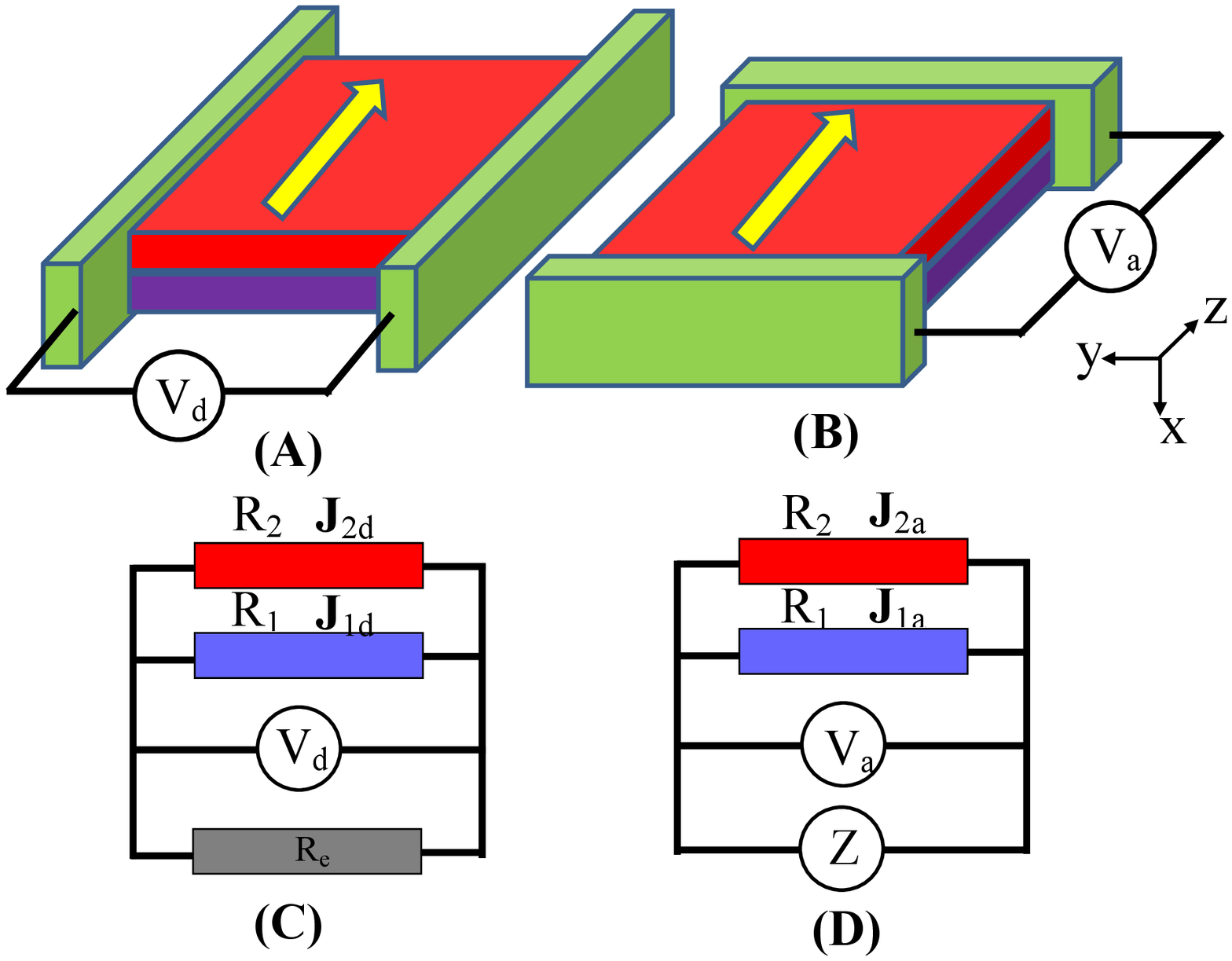}\caption{Schematic spin battery
operated by FMR, including the measurement configurations (A and B), and their
equivalent circuits (C and D). The AC(DC) voltage between both ends of the
sample drops along the $z$($y$) direction. $R_{1,2}$ in the equivalent circuit
represents the resistances of the normal metal and the ferromagnet,
respectively. $J_{1(2)d(a)}$ denote the DC (AC) charge current due to the ISHE
in the normal metal (1) or ferromagnet (2). $R_{e}$ is a parasitic resistance
in the external circuit for the DC, while $Z$ is the external impedance for
the AC measurements.}%
\label{fig1}%
\end{figure}

A normal metal N in contact with a ferromagnet F under FMR as shown in
Fig.~\ref{fig1} can be interpreted as a spin battery \cite{Brat02}.
When the ferromagnetic film is thicker than its transverse spin-coherence
length (a few monolayers), the adiabatically pumped spin current reads
\cite{Tser02a,Tser05,Gui11,Brat02}
\begin{equation}
\mathbf{I}_{s}^{\mathrm{p}}=\frac{\hbar}{4\pi}\left(
\mathrm{\operatorname{Re}}g^{\uparrow\downarrow}\mathbf{m}\times
\frac{d\mathbf{m}}{dt}+\operatorname{Im}g^{\uparrow\downarrow}\frac
{d\mathbf{m}}{dt}\right)  , \label{spp}%
\end{equation}
where $\mathbf{m}$ is the unit vector of the magnetization direction and
$g^{\uparrow\downarrow}$ is the (dimensionless) complex spin mixing
conductance \cite{Bra00}. The pumping spin current creates a spin accumulation
in N that induces a diffusion backflow of spins into F:
\begin{align}
\mathbf{I}_{s}^{\mathrm{b}}  &  =\frac{g}{8\pi}[2p(\mu_{0}^{F}-\mu_{0}%
^{N})+\mu_{s}^{\mathrm{F}}-\mathbf{m}\cdot\boldsymbol{\mu}_{s}^{N}%
]\mathbf{m}\nonumber\\
&  -\frac{\mathrm{\operatorname{Re}}g^{\uparrow\downarrow}}{4\pi}%
\mathbf{m}\times\left(  \boldsymbol{\mu}_{s}^{N}\times\mathbf{m}\right)
+\frac{\operatorname{Im}g^{\uparrow\downarrow}}{4\pi}\mathbf{m}\times
\boldsymbol{\mu}_{s}^{N}, \label{spb}%
\end{align}
where $\mu_{0}^{N}$, $\boldsymbol{\mu}_{s}^{N}$ in N and $\mu_{0}^{F}$,
$\mu_{s}^{\mathrm{F}}\mathbf{m}$ in F are the charge and spin accumulations at
the interface. The sum of spin-up and spin-down interface conductances is the
total conductance $g=g^{\uparrow\uparrow}+g^{\downarrow\downarrow}$ with
$p=\left(  g^{\uparrow\uparrow}-g^{\downarrow\downarrow}\right)  /\left(
g^{\uparrow\uparrow}+g^{\downarrow\downarrow}\right)  $ the conductance spin
polarization. The magnetization determined by the Landau-Lifshitz-Gilbert
equation is assumed to precess with constant cone angle $\theta$ around the
$z$-axis, whose magnitude is governed by the rf radiation intensity.
The spin accumulation in N obeys the spin-diffusion equation \cite{John88}
\begin{equation}
\frac{\partial\boldsymbol{\mu}_{s}^{N}(\mathbf{r},t)}{\partial t}=\gamma
_{N}\mathbf{H}_{ex}\times\boldsymbol{\mu}_{s}^{N}+D_{N}\frac{\partial
^{2}\boldsymbol{\mu}_{s}^{N}}{\partial x^{2}}-\frac{\boldsymbol{\mu}_{s}^{N}%
}{\tau_{sf}^{N}}, \label{diff}%
\end{equation}
where $\gamma_{N}$ is the gyromagnetic ratio,
$D_{N}$ the diffusion constant and $\tau_{sf}^{N}$ the spin-flip relaxation
time, all in N. The spin current $\mathbf{I}_{s}=\mathbf{I}_{s}^{\mathrm{p}%
}+\mathbf{I}_{s}^{\mathrm{b}}$ is continuous at the N$|$F interface and
vanishes at the outer boundary $x=d_{N}$.
In position-frequency space the exact solution for the spatiotemporal
dependence of the spin accumulation reads
\begin{equation}
\boldsymbol{\mu}_{s}^{N}(x,\omega)=\sum_{i=1}^{3}\vec{v}_{i}\frac
{\mathrm{\cosh}[\kappa_{i}(x-d_{N})]}{\mathrm{\sinh}[\kappa_{i}d_{N}]}%
\frac{2j_{s}^{i}(x=0,\omega)}{\hbar\nu D_{N}\kappa_{i}}.
\end{equation}
$\kappa_{1}^{2}(\omega)=(1+i\omega\tau_{sf}^{N})/(\lambda_{sd}^{N})^{2}$,
$\kappa_{2,3}^{2}(\omega)=\kappa_{1}^{2}(\omega)\pm iC$, $C=-\gamma_{N}%
H_{ex}/D_{N}$ and $\lambda_{sd}^{N}=\sqrt{D_{N}\tau_{sf}^{N}}$. $j_{1s}%
=I_{s}^{z}/A$ and $j_{\left(  2,3\right)  s}=(I_{xs}\pm iI_{ys})/\left(
\sqrt{2}A\right)  $ are spin current densities, where $A$ is the interface
area. Three eigenvectors associated with $\kappa_{i}^{2}(\omega)(i=1,2,3)$
are, respectively, $\vec{v}_{1}=%
\begin{pmatrix}
0 & 0 & 1
\end{pmatrix}
,\vec{v}_{2}=%
\begin{pmatrix}
1 & -i & 0
\end{pmatrix}
/\sqrt{2},\vec{v}_{3}=%
\begin{pmatrix}
1 & i & 0
\end{pmatrix}
/\sqrt{2}.$ In position-time domain
\begin{equation}
\mathbf{j}_{1s}(x,t)=-\frac{\hbar\nu D_{N}}{2}\frac{\partial\boldsymbol{\mu
}_{s}^{N}(x,t)}{\partial x}=j_{1s}^{z}(x)\mathbf{e}_{z}+\mathbf{j}_{1s}%
^{a}(x,t),
\end{equation}
with
\begin{align}
j_{1s}^{z}(x)\mathbf{e}_{z}  &  =\frac{\mathrm{\sinh}[\kappa_{1}(0)(d_{N}%
-x)]}{\mathrm{\sinh}[\kappa_{1}(0)d_{N}]}j_{1s}^{z}(0)\mathbf{e}_{z},\\
\mathbf{j}_{1s}^{a}(x)  &  =2\operatorname{Re}\left\{  \frac{\mathrm{\sinh
}[\kappa_{2}(\omega)(d_{N}-x)]}{\mathrm{\sinh}[\kappa_{2}(\omega)d_{N}%
]}\mathbf{j}_{1s}^{a}(0)e^{i\omega t}\right\}
\end{align}
$j_{1s}^{z}(0)$ and $\mathbf{j}_{1s}^{a}(0)$ are complicated analytic
expression (not shown) for the DC and AC components of the spin current at the
N-side of the interface.

The longitudinal component of the spin accumulation can penetrate into a
metallic ferromagnet, leading to a spin accumulation $\mathbf{m}(t)\mu
_{s}^{\mathrm{F}}$. $\mu_{s}^{\mathrm{F}}=\mu_{\uparrow}^{\mathrm{F}}%
-\mu_{\downarrow}^{\mathrm{F}}$ that satisfies the spin-diffusion
equation\textit{ }
\begin{equation}
\frac{\partial^{2}\mu_{s}^{\mathrm{F}}(x)}{\partial x^{2}}=\frac{\mu
_{s}^{\mathrm{F}}(x)}{(\lambda_{sd}^{\mathrm{F}})^{2}},
\end{equation}
where $\lambda_{sd}^{\mathrm{F}}$ is the spin-flip diffusion length in the
ferromagnet. In an open circuit the DC charge current vanishes and we obtain
\begin{equation}
\mu_{s}^{\mathrm{F}}(x)=\frac{\mathrm{\cosh}[(d_{F}+x)/\lambda_{sd}%
^{\mathrm{F}}]\tilde{g}}{[g_{F}\mathrm{\tanh}[d_{F}/\lambda_{sd}^{\mathrm{F}%
}]+\tilde{g}]\mathrm{\cosh}(d_{F}/\lambda_{sd}^{\mathrm{F}})}\mathbf{m}%
\cdot\boldsymbol{\mu}_{s}^{N},
\end{equation}
where $g_{F}=4hA\sigma_{\uparrow}\sigma_{\downarrow}/[e^{2}\lambda
_{sd}^{\mathrm{F}}(\sigma_{\uparrow}+\sigma_{\downarrow})]$ , $\tilde
{g}=(1-p^{2})g$. Here, $\sigma_{\uparrow(\downarrow)}$ is the conductivity of
spin-up(-down) electrons in F. The spin current density in F reads
\begin{equation}
\mathbf{j}_{2s}(x)=\frac{\mathrm{\sinh}[(d_{F}+x)/\lambda_{sd}^{F}%
]}{\mathrm{\sinh}\left(  d_{F}/\lambda_{sd}^{F}\right)  }\mathbf{j}_{2s}(0).
\end{equation}
At the interface
\begin{equation}
\mathbf{j}_{2s}(0)=-\frac{1}{8\pi}\frac{\tilde{g}g_{F}\mathrm{\tanh}%
[d_{F}/\lambda_{sd}^{\mathrm{F}}]}{\tilde{g}+g_{F}\mathrm{\tanh}[d_{F}%
/\lambda_{sd}^{\mathrm{F}}]}(\mathbf{m}\cdot\boldsymbol{\mu}_{s}%
^{N})\mathbf{m}.
\end{equation}
When spin-flip in F is negligible, $\lambda_{sd}^{\mathrm{F}}\gg d_{F}$,
$\mu_{s}^{\mathrm{F}}(0)=\mathbf{m}\cdot\boldsymbol{\mu}_{s}^{N}$ and the spin
current in F vanishes. The backflow spin current modifies the magnetization
dynamics by contributing a three-component transfer torque that (i) reduces
the interface Gilbert damping due to spin pumping, (ii) modulates the
gyromagnetic ratio, and (iii) adds an effective magnetic field. For the system
parameters considered below the last two terms are too small to affect the
magnetizations dynamics, however.

The ISHE generates a charge current transverse to an applied spin current due
to the spin-orbit interaction, which for in an open circuit generates electric
fields. With spin current along the $x$-direction, the charge current reads
\cite{Sait06,Azeve05,Czes11,Mose10a,Mose10b,Azeve11}
\begin{equation}
\mathbf{j}_{c}(x)=\alpha_{N/F}(2e/\hbar)\mathbf{e}_{x}\times\mathbf{j}_{s}(x),
\end{equation}
where $\alpha_{N}$ is the spin Hall angle in N and $\alpha_{F}=(\alpha
_{F\uparrow}+\alpha_{F\downarrow})/2$ is that in F, where $\alpha_{F\xi
}=\sigma_{AH\xi}/\sigma_{\xi}$ $(\xi=\uparrow,\downarrow)$ and $\sigma
_{\left(  AH\right)  \xi}$ is the spin-polarized (anomalous Hall) conductivity.

As shown in Fig. \ref{fig1}(A) (\ref{fig1}(B)), a DC (AC) component can be
detected along the $y$ $(z)$ direction by the electric fields $E_{y}%
\mathbf{e}_{y}$ $\left(  E_{z}(t)\mathbf{e}_{z}\right)  $. We consider the
equivalent circuits shown in Fig. \ref{fig1}(C) (\ref{fig1}(D)) and disregard
parasitic impedances, \textit{i.e}. assume that $R_{e},Z\rightarrow\infty$. In
the steady state, we obtain an AC electric field:
\begin{align}
E_{z}(t)  &  =\frac{4e/\hbar}{\sigma_{N}d_{N}+\sigma_{F}d_{F}}%
\operatorname{Re}\left(  \frac{\alpha_{N}j_{1s}^{y}\left(  0\right)  }%
{\kappa_{2}\left(  \omega\right)  }\tanh\frac{d_{N}\kappa_{2}\left(
\omega\right)  }{2}\right. \nonumber\\
&  \left.  +\alpha_{F}j_{2s}^{y}(0)\lambda_{sd}^{F}\tanh\frac{d_{N}}%
{2\lambda_{sd}^{N}}\right)  . \label{eq17}%
\end{align}
The DC electric field along the $y$-direction reads
\begin{align}
E_{y}=  &  \frac{2e/\hbar}{\sigma_{N}d_{N}+\sigma_{F}d_{F}}\left[  j_{1s}%
^{z}\left(  0\right)  \alpha_{N}\lambda_{sd}^{N}\tanh\frac{d_{N}}%
{2\lambda_{sd}^{N}}\right. \nonumber\\
&  \left.  +j_{2s}^{z}\left(  0\right)  \alpha_{F}\lambda_{sd}^{F}\tanh
\frac{d_{F}}{2\lambda_{sd}^{F}}\right]  . \label{eq18}%
\end{align}
These equations are our main results. In the following we disregard
$\operatorname{Im}g^{\uparrow\downarrow}$ which is small for the interfaces
considered below. When backflow is disregarded we recover the relation
\cite{Mose10a,Mose10b}
\begin{equation}
E_{y}^{NB}=\frac{e\alpha_{N}f_{FMR}\sin^{2}\theta}{\sigma_{N}d_{N}+\sigma
_{F}d_{F}}\frac{\mathrm{\operatorname{Re}}g^{\uparrow\downarrow}}{A}%
\lambda_{sd}^{N}\tanh\frac{d_{N}}{2\lambda_{sd}^{N}}.
\end{equation}
as well as the AC signal
\begin{equation}
\frac{E_{z}^{NB}(t)}{\cos(\omega t+\delta)}=\frac{e\alpha_{N}f_{FMR}\sin
\theta}{\sigma_{N}d_{N}+\sigma_{F}d_{F}}\frac{\mathrm{\operatorname{Re}%
}g^{\uparrow\downarrow}}{A}\left\vert \frac{\tanh[\kappa_{2}(\omega)d_{N}%
/2]}{\kappa_{2}(\omega)}\right\vert ,
\end{equation}
where $\delta=\delta_{0}+\operatorname{Arg}\{\tanh[\kappa_{2}(\omega
)d_{N}/2]/\kappa_{2}(\omega)\}$, with $\delta_{0}=-\pi$ for $\alpha_{N}>0$ and
$0$ for $\alpha_{N}<0$, is the phase shift relative to the ac excitation field
$\cos(\omega t)\mathbf{e}_{y}$.

The spin-pumping induced spin accumulation is governed by two length scales,
$\lambda_{sd}^{N}$ and the transverse spin dephasing length $\lambda_{c}%
=\sqrt{D_{N}/\omega}$ \cite{Tser05}. For $x\ll\lambda_{c}$ the spin
accumulation follows rigidly the instantaneous $\mathbf{m}\times
\mathbf{\dot{m}}$ polarization and contributions to both AC and DC ISHE are
large. When $x\gg\lambda_{c}$ the transverse component is dephased such that
only a time-independent $\mathbf{e}_{z}$ component remains \cite{Brat02} that
contributes to the DC, but not the AC signal. From $\omega\ll1/\tau_{sf}^{N}$
follows $\lambda_{sd}^{N}\ll\lambda_{c}$. This is the case when $\omega
\tau_{sf}^{N}\simeq0.2\left(  f_{FMR}/10\operatorname{GHz}\right)  \left(
\tau_{sf}^{N}/3\operatorname{ps}\right)  \ll1$. $\omega\tau_{sf}^{N}%
=1\times10^{-3}$ in Pt and $1.5\times10^{-2}$ in Ta at $f_{FMR}%
=15.5\operatorname{GHz}$ with $\tau_{sf}^{Pt}=0.01\operatorname{ps}$ and
$\tau_{sf}^{Ta}=0.15\operatorname{ps}$ calculated from the data in Table
\ref{tb4}.
So in the present frequency region this condition is fulfilled for strongly
spin-dissipating metals. The diffusive dephasing of the time-dependent spin
accumulation is then small and the AC\ ISHE is maximal. In that limit the
contribution from the anomalous Hall effect in a ferromagnet such as Py is
found to be negligible. Previous expressions for the DC spin accumulation
\cite{Tser05} and voltages \cite{Naka12,Castel} agree with the present results
in that limit. Note that this condition does not hold for metals with long
spin-flip times for example single crystal Al.

\begin{figure}[ptb]
\includegraphics[width=8cm]{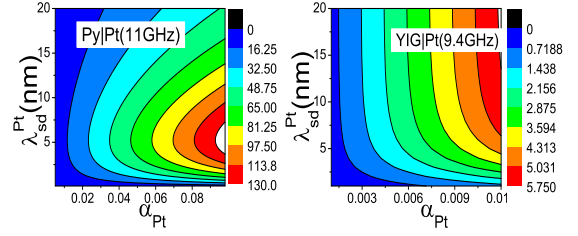}\caption{The calculated DC field (in
units of $\operatorname{\mu V}/\operatorname{mm}$) as a function of the spin
Hall angle and the spin-diffusion length (with backflow).}%
\label{figdc}%
\end{figure}

\begin{figure}[ptb]
\includegraphics[width=6cm]{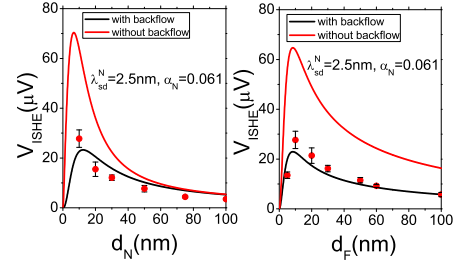}\caption{Theoretical and experimental
\cite{Naka12} ISHE voltages in Pt$|$Py bilayers as a function of Pt thickness
(A) and Py thickness (B). }%
\label{fig2}%
\end{figure}

In Fig. \ref{figdc}, we plot the DC electric fields with backflow of spin as a
function of spin Hall angle $\alpha_{N}$ and spin diffusion length
$\lambda_{sd}^{N}$ (noting that the results are very insensitive to changes in
$\lambda_{sd}^{F}$ and $\alpha_{F}$). The strong correlation between these two
parameters especially for YIG$|$Pt is evident. Nevertheless, we can narrow
them down when also the Gilbert damping enhancement is measured, as was done
recently by Nakayama\textit{ et al.} \cite{Naka12}. The spin-mixing
conductance $(\mathrm{\operatorname{Re}}g_{\mathrm{eff}}^{\uparrow\downarrow
}/A)^{-1}=(\mathrm{\operatorname{Re}}g^{\uparrow\downarrow}/A)^{-1}%
+\Gamma^{-1}=3\times10^{19}%
\operatorname{m}%
^{-2}$ where $\Gamma=(h/e^{2})\sigma_{N}/\lambda_{sd}^{N}$ for $d_{N}%
\gg\lambda_{sd}^{N}\ $can be obtained from the Gilbert damping constant
$\zeta_{\mathrm{eff}}=\zeta_{0}+\left(  g\mu_{B}\right)  /\left(  4\pi
M_{s}d_{F}\right)  \mathrm{\operatorname{Re}}g_{\mathrm{eff}}^{\uparrow
\downarrow}/A$. The conductances are parameterized as $\sigma_{N}%
=4.1\times(1-e^{-d_{N}/29.6})10^{6}%
\operatorname{\Omega }%
^{-1}%
\operatorname{m}%
^{-1}$ and $\sigma_{F}=3.5\times(1-e^{-d_{F}/9.8})10^{6}%
\operatorname{\Omega }%
^{-1}%
\operatorname{m}%
^{-1}$ to fit the experiments (H. Nakayama, private communication). Since
$\mathrm{\operatorname{Re}}g^{\uparrow\downarrow}>0$ the experiments provide
the important constraint that $\lambda_{sd}^{N}<3.5\,%
\operatorname{nm}%
,$ which is not consistent with larger values in use for this parameter. The
constraint that the spin-flip scattering relaxation time should be larger than
the scattering lifetime leads to $\lambda_{sd}^{N}>2\,%
\operatorname{nm}%
$.
In Fig. \ref{fig2} we plot the computed and the measured spin Hall voltages as
a function of the layer thicknesses for optimized parameter combinations with
and without backflow. The largest ISHE voltages are generated for $d_{N}%
\simeq\lambda_{sd}^{N}$ (see below)\cite{Azeve11,Naka12,Castel}. Since above
estimates favor $\lambda_{sd}^{N}\approx2.5\,%
\operatorname{nm}%
$, we estimate the Hall angle $\alpha_{Pt}\approx0.061\ $from the spin pumping
experiments \cite{Naka12} and consistency arguments alone. These parameters
are different from those reported\cite{Azeve11,Naka12}, illustrating the
importance of taking into account the experimental constraint on $\lambda
_{sd}^{N}$ provided by the increased Gilbert damping. 
\begin{table}[ptb]
\caption{Parameters for selected bilayer systems used to compute AC ISHE
electric fields induced by spin pumping under FMR. }%
\label{tb4}%
\begin{tabular}
[c]{ccccccl}\hline\hline
N & $^{a}\nu_{DOS}$ & $\sigma_{N}$ & $\lambda_{sd}^{N}$ & $\alpha_{N}$ &
$g_{N}^{sh}$ & \\
& [$10^{47}J^{-1}m^{-3}$] & [$10^{6}\Omega^{-1}m^{-1}$] & [nm] &  &
[$10^{19}m^{-2}$] & \\\hline
Al & 1.5 & $^{b}$11 & $^{b}350$ & $^{c}$0.0001 & 3.6 & \\
Ta & 4.3 & $^{d}$0.53 & $^{e}$2.7 & $^{d}$-0.15 & 2.5 & \\
Au & 1.1 & $^{f}$25.2 & $^{f}$35 & $^{f}$0.0035 & 1.2 & \\
Pd & 10.0 & $^{f}$4.0 & $^{f}$15 & $^{f}$0.0064 & 1.6 & \\
Pt & 9.1 & $^{g}$5 & $^{g}$1.5 & $^{g}$0.07 & 1.8 & \\\hline\hline
& Py$|$NM & YIG$|$Au &  & YIG$|$Pt & p & \\
$\frac{\operatorname{Re}g^{\uparrow\downarrow}}{A}$ & $2g_{N}^{sh}$ &
$^{h}0.66$ &  & 2.3 & 0.4 & \\\hline
\end{tabular}
\begin{tabular}
[c]{l}%
$^{a}$Ref. \onlinecite{Papa86},$^{b}$Ref. \onlinecite{Jed02},$^{c}$Ref.
\onlinecite{Valen06},$^{d}$Ref. \onlinecite{Liu12b},$^{e}$Ref.
\onlinecite{Moro11},$^{f}$Ref.\onlinecite{Mose10a},$^{g}$%
Ref.\onlinecite{Liu12}\\
$^{h}$Ref. \onlinecite{Burr11}. Schep corrections\cite{Schep97,Bauer02} are
included in $\operatorname{Re}g^{\uparrow\downarrow}/A$.\\\hline\hline
\end{tabular}
\end{table}

\begin{figure}[ptb]
\includegraphics[width=8cm]{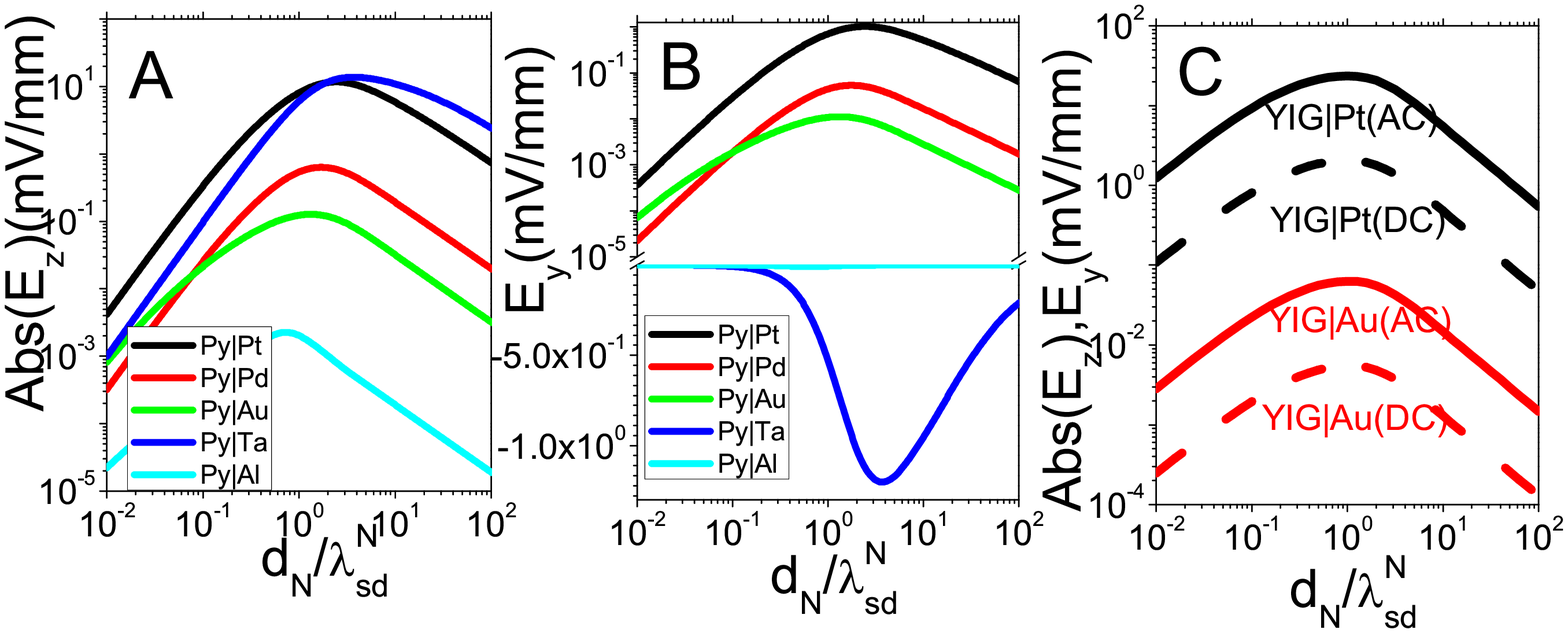}\caption{ The AC and DC electric fields
as a function of $d_{N}$ for Py$|$N and YIG$|$N for a fixed FMR frequency of
$15.5\,\operatorname{GHz}$. Here, the precession angle is $5^{\circ}$ and
$d_{F}=15\,\operatorname{nm}$. }%
\label{fig4}%
\end{figure}

In Fig.\ref{fig4} we turn to the AC-ISHE by comparing its dependence on the
normal metal thickness with the DC counterpart for a precession angle of
$5^{\circ}$ for Py$|$N (N=Au, Ta, Pd, Pt and Al) and YIG$|$N (N=Pt,Au)
bilayers. In Py, we choose the electrical conductivity $\sigma_{Py}%
=1.5\times10^{6}%
\operatorname{\Omega }%
^{-1}%
\operatorname{m}%
^{-1}$ and the conductivity polarization $q=(\sigma_{Py}^{\uparrow}%
-\sigma_{Py}^{\downarrow})/\sigma_{Py}=0.7$, estimate the spin Hall angle
$\alpha_{Py}=0.076$ from the Hall electrical conductivity $\sigma
_{H}=0.09\times10^{6}\mathrm{\Omega}^{-1}\mathrm{m}^{-1}$\cite{Miya07} and its
polarization $p_{H}=(\sigma_{H\uparrow}-\sigma_{H\downarrow})/\sigma_{H}=0.5$.
Its spin-diffusion length is chosen as $\lambda_{sd}^{F}=5%
\operatorname{nm}%
$ \cite{Bass03}. Both ISHE fields are maximized for $d_{N}\sim\lambda_{sd}%
^{N}$ since the AC and DC signals are affected by two factors, \textit{i.e}.
the effective spin current and the effective resistance. Increasing $d_{N}$
from zero, the total spin current initially increases exponentially because of
the reduced backflow. When the thickness increases further, the emf generated
by the ISHE close to the interface is short-circuited by the normal metal
region $x\gtrsim\lambda_{sd}^{N},$ leading to an algebraic decrease of the
voltage for larger $d_{N}$. A systematic experimental study of the DC ISHE as
a function of $d_{N}$ around the spin-diffusion length should help to
understand the backflow and lead to a more accurate parameter determinations,
including $\lambda_{sd}^{N}$. 

\begin{figure}[ptb]
\includegraphics[width=4.0cm]{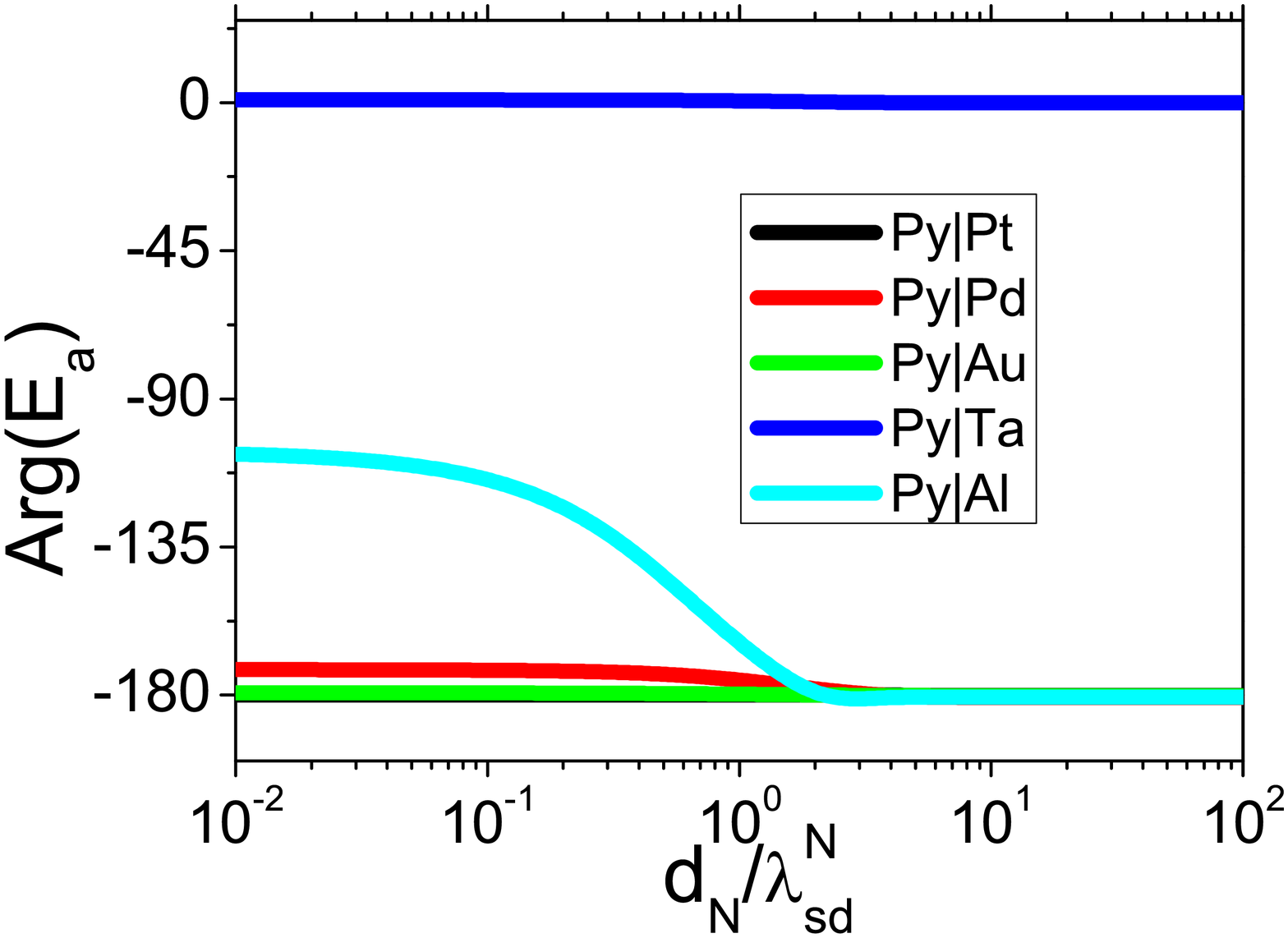}
\includegraphics[width=4.0cm]{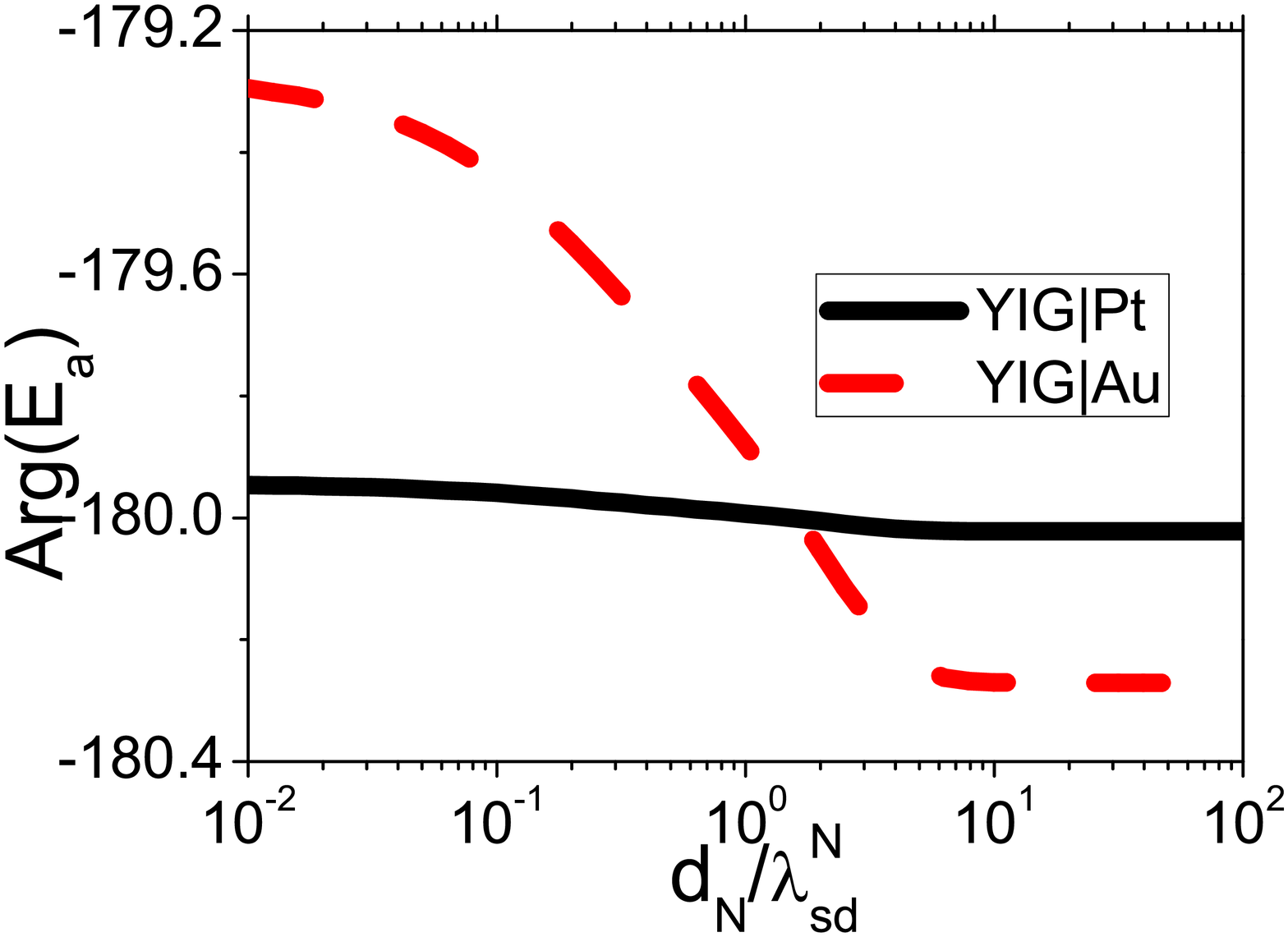}\caption{Phase shift of the AC-ISHE
relative to the rf magnetic field as function of $d_{N}$ (with backflow).}%
\label{figa}%
\end{figure}

We find that the anomalous Hall effect in Py caused by the backflow of spins
into the ferromagnet is negligible unless the ISHE in the normal metal is very
small, as \textit{e.g}. in single crystal Al. In Py$|$N or YIG$|$N the phase
shifts depend only very weakly on $d_{N}$, again with the exception for a
light metal such as Al. In Fig.\ref{figa} we plot the phase of the AC-ISHE for
selected bilayers. The phase difference of $180^{\circ}$ at large $d_{N}$ is
simply caused by the different sign of the spin Hall angles for Pt(Pd,Au) and
Ta. Again, interesting effects can be observed for a material with little spin
dissipation such as Al, in which the phase is affectd by the anomalous Hall
effect in Py.
For constant precession angles, the voltages increases with FMR frequency due
to the increased spin pumping $\sim\left\vert \mathbf{\dot{m}}\right\vert $.
When the rf intensity is kept constant with frequency, the precession angle is
inversely proportional to the FMR frequency. By increasing $\omega$ for small
cone angles the DC voltage therefore decreases, while the AC voltage is almost
unchanged .

The AC voltage is proportional to the precession angle, or the square root of
the AC excitation power, in contrast to the linear relation between DC voltage
and excitation power \cite{Mose10a,Mose10b,Ando11}. Furthermore, the ratio of
the AC to DC field moduli
is much larger than unity for the intensities typical for FMR experiments.
This ratio is close to a universal function as long as the anomalous Hall
effect does not play a role (always the case for magnetic insulators)
approaching the scaling function $C(\omega)\cot\theta$, where $C(\omega)$ is
material-dependent. When $\omega\ll1/\tau_{sf}^{N}$, $C(\omega)\simeq1$, which
is the case for Pt, Pd, Au and Ta.

In summary, we present a dynamical theory of the ISHE detection of spin
pumping, explicitly including the back-diffusion of spins into the
ferromagnet. We predict the generation of an AC voltage along the effective
magnetic field in F$|$N bilayers under FMR. We predict magnitudes and phase
shifts of the AC voltages for Py$|$N and YIG$|$N. From the analysis of
published experiments, we predict that the spin Hall angle in Pt is $0.06$. If
the ISHE signal can be separated from parasitic voltages at the resonance
frequency, the larger signals of AC measurements could be an attractive
alternative to detect spin currents.

This work was supported by the FOM Foundation, EU-ICT-7 \textquotedblleft
MACALO\textquotedblright, the ICC-IMR, and DFG Priority Programme 1538
\textquotedblleft Spin-Caloric Transport\textquotedblright. We thank Profs.
Can-Ming Hu, Bechara Muniz, Sergio Rezende, and MinZhong Wu for their comments
on the first version of the manuscript.

\end{document}